\documentstyle[preprint,revtex]{aps}
\begin{document}
\draft
\preprint{IFT-P.039/92}
\preprint{December 1992}
\begin{title}
Neutral currents and GIM mechanism\\ in $SU(3)_L\otimes U(1)_N$ models
for electroweak interactions
\end{title}
\author{ J. C. Montero, F. Pisano and V. Pleitez}
\begin{instit}
Instituto de F\'\i sica Te\'orica \\
Universidade Estadual Paulista \\
Rua Pamplona, 145 \\
01405-900--S\~ao Paulo, SP \\
Brazil
\end{instit}
\begin{abstract}
We study the Glashow-Iliopoulos-Maiani mechanism for
flavor-changing-neutral-currents suppression in both, the gauge and
Higgs sectors, for models with $SU(3)_L\otimes U(1)_N$ gauge symmetry.
The models differ one from the other only with respect to the
representation content. The main features of these models are that in
order to cancel the triangle anomalies the number of families must be
divisible by three (the number of colors) and that the lepton number
is violated by some lepton-gauge bosons and lepton-scalar
interactions.
\end{abstract}
\section{Introduction}
\label{sec:intro}
It is a well known fact that flavor changing neutral currents (FCNC)
are very suppressed with respect to the charged currents weak
interactions. This follows from experimental data in decays as
$K^0_L\to \mu^+\mu^-$, $D^0\to\mu^+\mu^-$ and $B^0\to \mu^+\mu^-$ for
transitions $s\leftrightarrow d$, $c\leftrightarrow u$ and
$b\leftrightarrow d$ respectively. Experimentally it is
found that~\cite{pdg}
\begin{equation}
\Gamma(K_L\to \mu^-\mu^-)/\Gamma(K_L\to
\mbox{all})=(7.3\pm0.4)\times 10^{-9},
\label{exp1}
\end{equation}
\begin{equation}
\Gamma(D^0\to\mu^+\mu^-)/\Gamma(D^0\to\mbox{all})<1.1\times 10^{-5},
\label{exp2}
\end{equation}
\begin{equation}
\Gamma(B^0\to\mu^+\mu^-)/\Gamma(B^0\to\mbox{all})<1.2\times 10^{-5}.
\label{exp3}
\end{equation}
Change of flavor by two units as $\vert\Delta S\vert=2$ and
$\vert\Delta B\vert=2$ are very suppressed too. In the Standard Model
this occurs only in second-order weak interactions. The two examples
for which this have been measured are the $K^0-\bar K^0$ and $B^0-\bar
B^0$ mass difference~\cite{pdg}
\begin{equation}
m_{K_S}-m_{K_L}=(3.522\pm0.016)\times 10^{-12}\,\mbox{MeV},
\label{exp4}
\end{equation}
\begin{equation}
\vert m_{B^0_1}-m_{B^0_2}\vert=(3.6\pm0.7)\times 10^{-10}\,\mbox{MeV},
\label{exp5}
\end{equation}
On the other hand, no evidence exists for $D^0-\bar D^0$ mixing.
Next, from the experi\-mental data we infer that FCNC effects in the
$s-d$, $c-u$ and $d-b$ systems are smaller than $O(\alpha G_F)$, but we
do not know if the same occurs in the $s-b$ sector and in other
systems involving the, so long not discovery, top quark.

In the Standard Model, which is based on the $SU(2)_L\otimes U(1)_Y$
gauge symmetry~\cite{ws}, the Glashow-Iliopoulos-Maiani (GIM)
mechanism accounts for the suppression of neutral processes in which
there is a change of flavor in order $G_F$ or $\alpha G_F$~\cite{gim}.

The problem of how to implement such a suppression of FCNC effects in
$SU(2)_L\otimes U(1)_Y$ models in a {\em natural} way was considered
a long time ago by Glashow and Weinberg~\cite{gw}. The term {\em
natural} means that the conservation of flavor in neutral currents
follows from the group structure and representation content of the
theory, that is, the suppression of FCNC is valid for all values of
the parameters of the theory. The necessary and sufficient conditions
are, that all quarks of fixed charge and helicity must (i) transform
according to the same irreducible representation of the $SU(2)$
group, (ii) have the same values of weak $I_3$, and (iii) gain mass
from a single source.

As the main points are valid for a general kind of models we revised
them briefly. The neutral currents have in general the form~\cite{gw}
\begin{equation}
J^\mu_Z=\bar q_L\gamma^\mu Y_Lq_L+\bar q_R\gamma^\mu Y_Rq_R,
\label{nc}
\end{equation}
where $q$ denotes a column vector with all quark fields
(phenomenological states) and $Y_L,Y_R$
are the matrices
\[Y_L\propto I_{3L}-2\sin^2\theta_W Q,\]
\begin{equation}
Y_R\propto I_{3R}-2\sin^2\theta_W Q,
\label{y}
\end{equation}
where $\theta_W$ is the electroweak mixing angle, $I_{3L}(I_{3R})$
are the matrices corresponding to the third component of the
$SU(2)$ group for the left-handed (right-handed) fields,
and $Q$ is the charge operator. On the other hand, the mass term of the
Lagrangian have the form
\begin{equation}
-\bar q_LM^Qq_R-\bar q_RM^{Q\dagger} q_L,
\label{m}
\end{equation}
where the mass matrix for the charge-$Q$ sector, $M^Q$, is in general,
neither Hermitian nor diagonal. It is possible to redefine the quark
fields as
\begin{equation}
Q_L=V^Q_Lq_L\quad Q_R=V^Q_Rq_R,
\label{def}
\end{equation}
with $V^Q_{L},V^Q_{R}$ unitary matrices in the flavor
space. In the $Q_{L,R}$ basis
\begin{equation}
M'^Q=V^Q_LM^Q{V_R^Q}^{-1},
\label{m'}
\end{equation}
with $M'^Q$ diagonal. The physical states ($Q_{L,R}$ fields) are
defined as eigenstates of the quark mass matrix.

In order to have natural suppression of FCNC it is necessary that in
Eq.~(\ref{y}) all quarks with the same charge have the same
values of the third component of $I_L$ and $I_R$, which
implies that these components are functions of the electric
charge~\cite{gw},
\begin{equation}
I_{3L}=f_L(Q),\qquad I_{3R}=f_R(Q).
\label{i}
\end{equation}
The conditions (\ref{i}) imply the GIM mechanism in each charge
sector. In $SU(2)_L\otimes U(1)_Y$ gauge theories $\alpha G$ neutral
couplings induced at the one-loop level are suppressed if all quarks
with the same charge have the same value of $\vec I^2_L$ and $\vec
I^2_R$.

Later Georgi and Pais~\cite{gp} have generalized the GIM mechanism to
systems of more than four quarks and leptons in a different way. They
have expanded the quark and lepton systems in the ``vertical''
direction by enlarging the gauge group to $SU(3)\otimes U(1)$. They
called ``horizontal mixings'' the particle mixtures at equivalent
position within equivalent representation of the gauge group,
otherwise the particle mixtures are called ``vertical mixings''.

In this work we will consider the group $SU(3)_L\otimes
U(1)_N$~\cite{pp,vs,svs} as
the gauge symmetry with several representation contents.
Some FCNC are naturally suppressed in the horizontal mixings in the
gauge and in the Higgs sectors, but vertical mixings are associate
with FCNC. This happens in the Higgs sector because, quarks with the
same charge but in non-equivalent representations, gain mass from
different sources.

In all these models there are two real neutral bosons, however currents
coupled to the lightest $Z^0$ neutral boson implement the GIM
mechanism naturally. Notwithstanding, FCNC are induced by the heaviest
neutral boson $Z'^0$ (which we will call ``Zezeon''), but as it gains
mass from the Higgs field which breaks the $SU(3)_L\otimes U(1)_N$
symmetry into the $SU(2)_L\otimes U(1)_Y$ one, its mass is arbitrarily large.

In this kind of models the lepton number is a gauge
symmetry which is spontaneously broken in the interactions of leptons
with gauge and scalar bosons. This is so because we have both,
particles and antiparticles in the same multiplets~\cite{pp,vs,svs}.

The organization of this paper is as follows. In Sec.~\ref{sec:model}
we consider three representation contents for the same gauge symmetry.
In Sec.~\ref{sec:yukawa} we give the Yukawa Lagrangian
for the matter field representations given in the previous section.
The vector boson sector for one of the model is presented in details
in Sec.~\ref{sec:gb}. The charged and neutral currents are given in
Sec.~\ref{sec:currents}. Sec.~\ref{sec:gim} is devoted to show
explicitly that the GIM mechanism is in fact implemented because
quarks of the same charge have the same coupling with the $Z^0$
neutral boson. In one of the models (model I) it is necessary to add
some discrete symmetries which ensure that the Higgs fields give a
quark mass matrix of the tensor product form. Our conclusion are
summarized in the last section. In the Appendix we give the
$\bar\lambda$'s ($\lambda$'s) matrices for the antitriplet (triplet)
representation that we have used in this work.

\section{$SU(3)_L\otimes U(1)_N$ models}
\label{sec:model}
As we have mentioned in the last section, we will consider an
$SU(3)_L\otimes U(1)_N$ gauge theory for the weak interactions of
quarks and leptons. We assume that the strong interactions are
described by the unbroken color $SU(3)_c$ but we have suppressed the
color indices. Notwithstanding, they are considered in order to
cancel anomalies. In fact, the sort of models we will consider have
the interesting feature that the number of family must be divisible
by the number of color, 3 in the present case, in order to make the
model anomaly free. This happens only if we have equal number of
triplets and antitriplets taking into account the color degree of
freedom and requiring the sum of all fermions charges to
vanish~\cite{pp,vs,svs}. This means that one of the quark multiplets
transforms identically to the leptons under $SU(3)_L\otimes U(1)_N$
and the other two quark multiplets transform similarly to each other
but differently to the leptons. As we will show, however, the
phenomenology does not depend on the choice of which quark
multiplet transforms in the same representation of the leptons.

This sort of models have became an interesting
possibility for an extension of the Standard Model since the LEP
measurements~\cite{pdg,jn} of the $Z^0$ width indicated that the
number of sequential families is three, and this feature has no
explanation within the Standard Model, otherwise these models are
undistinguishable from the Standard Model up to the currently  energy
achieved experimentally.
\subsection{Model I}
\label{subsec:m1}
This is a model with leptons,
$\nu^c_l,\nu_l,\,l=e,\mu,\tau$ and nine quarks, four with charge
$\frac{2}{3}$ and five with charge $-\frac{1}{3}$. Let us start by
defining the electric charge operator as
\begin{equation}
\frac{Q}{e}=\frac{\lambda_3}{2}+\frac{1}{\sqrt3}\frac{\lambda_8}{2}+N.
\label{oc1}
\end{equation}
The left-handed leptons are assumed to belong to the following
antitriplets, $({\bf3}^*,-\frac{1}{3})$~\cite{fn}:
\begin{equation}
\psi_{lL}=\left(\begin{array}{c}
\nu^c_l \\ \nu_l \\ l^-
\end{array}\right)_L ,
\label{l1}
\end{equation}
where $-\frac{1}{3}$ denotes the $U_N(1)$ quantum number, and
$\l=e,\mu,\tau$.

In the quark sector, the first and second ``generations'' are in triplets
$({\bf3},0)$, and the third one in antitriplet $({\bf3}^*,+\frac{1}{3})$:
\begin{equation}
\begin{array}{cc}
Q_{iL}=\left(\begin{array}{c}
u_i\\ d_i \\ d'_i
\end{array}\right)_L \!\!\sim ({\bf3},0),\,\, i=1,2;&\quad
Q_{3L}=\left(\begin{array}{c}
u_3 \\ u_4 \\ d_3
\end{array}\right)_L\sim({\bf3}^*,+\frac{1}{3})
\end{array}
\label{q1}
\end{equation}
 All right-handed charged fermions are taken to
be $SU(3)_L$ singlets. The representations above are symmetry
eigenstates and they are related to the mass eigenstates by
Cabibbo-like angles as we will discuss in Sec.~\ref{sec:yukawa}.
In fact it does not matter which generation transforms similarly
to the leptons because the three generations are well defined only
after the symmetry breaking.

We will introduce the following Higgs multiplets:
\begin{equation}
\begin{array}{lll}
\eta=\left(\begin{array}{l}
\eta^0 \\ \eta_1^- \\ \eta_2^-
\end{array}\right)\!\!\sim ({\bf3},-\frac{2}{3});\;&
\rho=\left(\begin{array}{c}
\rho^+ \\ \rho_1^0 \\ \rho_2^0
\end{array}\right),\, &
\sigma=\left(\begin{array}{c}
\sigma^+ \\ \sigma_1^0 \\ \sigma_2^0\end{array}\right)\sim
({\bf3},+\frac{1}{3}).
\end{array}
\label{h1}
\end{equation}
In order to avoid additional FCNC effects we will also assume that
the vacuum expectation value (VEV) of the $\rho$ and $\sigma$ fields
lie in the second and third component respectively,
\begin{equation}
\langle\rho^0_1\rangle\not=0,\;\langle\rho^0_2\rangle=0;\;
\langle\sigma^0_1\rangle=0,\;\langle\sigma^0_2\rangle\not=0\,.
\label{vev}
\end{equation}
Let us briefly comment about the question if the choice of VEV's in
Eq.~(\ref{vev}) is a {\em natural} one. We can see this by
considering the more general gauge invariant potential involving the
Higgs triplets $\eta,\sigma,\rho$ is
\begin{eqnarray}
V(\eta,\sigma,\rho) & = & \lambda_1(\eta^\dagger\eta-v^2_\eta)^2+
\lambda_2(\rho^\dagger\rho-v^2_\rho)^2
+\lambda_3(\sigma^\dagger\sigma-v^2_\sigma)^2\nonumber \\ & & \mbox{}
+\lambda_4[(\eta^\dagger
\eta-v^2_\eta)+(\rho^\dagger\rho-v^2_\rho)+(\sigma^\dagger\sigma-v^2_\sigma)]^2
\nonumber \\ & & \mbox{}+\lambda_5[(\rho^\dagger\rho)(\sigma^\dagger\sigma)-
(\rho^\dagger\sigma)(\sigma^\dagger\rho)]
+\lambda_6(\rho^\dagger\eta)(\eta^\dagger\rho)
\nonumber \\ & & \mbox{}
+\lambda_7(\sigma^\dagger\eta)(\eta^\dagger\sigma)
+\lambda_8(\rho^\dagger\sigma)^2+
f\epsilon^{ijk}\eta_i\rho_j\sigma_k +H.c.
\label{p}
\end{eqnarray}
where $\lambda_i>0,\,i=1,...6$, and $\epsilon^{ijk}$ is the totally
antisymmetric symbol. Notice that in Eq.~(\ref{p}) there are linear
terms in $\sigma^0_1$ and $\rho^0_2$. These terms and the Yukawa
interactions in Eqs.~(\ref{ly1}) and (\ref{ly11}) will induce
divergent loop corrections and a counter term will be
necessary, hence it would be impossible to maintain
$\langle\sigma^0_1\rangle=\langle\rho^0_2\rangle=0$
However, the trilinear terms $f$ are forbidden by appropriate discrete
symmetries  (see Eq.~\ref{ds} below) and the choice of the VEV's in
(\ref{vev}) is in fact a natural one.

\subsection{Model II}
\label{subsec:m2}
This model has already been studied in detail in Refs.~\cite{pp,fhpp}
hence, we will revisit it briefly here. In this model we define
the charge operator as follows
\begin{equation}
{Q\over e}= \frac{1}{2}\left(\lambda_3 -\sqrt{3}\lambda_8\right)+N.
\label{oc2}
\end{equation}
The leptons are in triplets $({\bf3},0)$
\begin{equation}
\psi_{lL}=\left(\begin{array}{c}
\nu_l \\ l \\ l^c\end{array}\right)_L
\label{l2}
\end{equation}
with $l=e,\mu,\tau$. The quarks belong to one triplet and two antitriplets,
\begin{equation}
\begin{array}{ll}
Q_{1L}=\left(\begin{array}{c}
 u_1 \\ d_1 \\ J_1
\end{array}\right)_L\!\!\sim ({\bf3},+\frac{2}{3});\;&
Q_{iL}=\left(\begin{array}{c}
J_i\\ u_i \\ d_i
\end{array}\right)_L \sim ({\bf3}^*,-\frac{1}{3}),\,i=2,3.
\end{array}
\label{q2}
\end{equation}
All right-handed fields are in singlets and the introduction of
right-handed neutrinos is optional. Notice that we have to introduce
exotic quarks of charge $\frac{5}{3}$ and $-\frac{4}{3}$, which we will call
``Josions''. The model also has exotic gauge bosons $Y^+,U^{++}$
which are very massive and we will call them ``Wanions'', and also
this is the case for the neutral Zezeon, $Z'^0$.

In this model it is necessary to
introduce three Higgs triplets $({\bf3},0)$, $({\bf3},-1)$ and
$({\bf3},1)$ in order to give mass to all the quarks.
Charged leptons obtain mass when we introduce a sextet
$({\bf6},0)$~\cite{fhpp}.

One of the Higgs, the one transforming as $({\bf3},-1)$, is
responsible for the first symmetry breaking
$SU(3)_L\otimes U(1)_N\to SU(2)_L\otimes U(1)_Y$.
As the Josions, Wanions and Zezeon obtain a contribution to their
masses from this triplet, they must be very heavy. The Josions masses
are in fact
arbitrary but Wanions and Zezeon must have mass larger than 4 TeV and
40 TeV respectively~\cite{pp}.
\subsection{Model III}
\label{subsec:m3}
This model is analogous to the previous one, but the leptons have
heavy charged partners,
\begin{equation}
\begin{array}{lll}
\left(\begin{array}{cl}
\nu_l \\ e^- \\ E^+\end{array}\right)_L;&
\left(\begin{array}{cl}
\nu_l \\ \mu^- \\ M^+\end{array}\right)_L;&
\left(\begin{array}{cl}
\nu_l \\ \tau^- \\ T^+\end{array}\right)_L,
\end{array}
\label{rmii}
\end{equation}
and now we have to introduce right-handed singlets for
$e^-_R,\mu^-_R,\tau^-_R$ and $E^+_R,M^+_R,T^+_R$. Only the Higgs
triplets of model II are required. The quark sector is the same of
model II and, the introduction of right-handed neutrinos is also
optional.
\section{Yukawa Lagrangian}
\label{sec:yukawa}
\subsection{Model I}
\label{subsec:m1y}
The Yukawa Lagrangian in terms of the symmetry eigenstates (we have
suppressed $SU(3)_L$ indices) is
\begin{eqnarray}
-{\cal L}_Y&=&G_{lm}\bar \psi_{lL} R_{m}\eta^*+\bar
Q_{i L}G^u_{i\alpha}U_{\alpha R}\eta +\bar
Q_{3L}(G^{t}_{3\alpha}\rho^*+G'^t_{3\alpha}\sigma^*)U_{\alpha R}\nonumber\\
&&\mbox{} +\bar
Q_{iL}(G^{d}_{i\beta}\rho+G'^d_{i\beta}\sigma)
D_{\beta R}+\bar Q_{3L}G^b_{3\beta}D_{\beta R}\eta^*+H.c.,
\label{yu1}
\end{eqnarray}
where $i=1,2$, $\alpha=1,2,3,4$, $\beta=1,2,3,4,5$ and
$U_{1,2,3,4 R}=u_{1R},u_{2R},u_{3R},u_{4R}$, $D_{1,2,3,4,5 R}
=d_{1R},d_{2R},d_{3R},d'_{1R},d'_{2R}$. For the leptons
$R_m=e_R,\mu_R,\tau_R$. Summation is assumed in the repeated indices.

{}From Eq.~(\ref{yu1}) we obtain the following mass terms
\begin{equation}
-\bar U_{\alpha L}M^U_{\alpha\alpha'}U_{\alpha' R}-\bar
D_{\beta L}M^D_{\beta\beta'}D_{\beta' R} + H.c.,
\label{mass1}
\end{equation}
where the indices $(\alpha,\alpha')$ and $(\beta,\beta')$, run over
all quarks with charge $+\frac{2}{3}$ and $-\frac{1}{3}$
respectively. Notice that the $M^U$ is a $4\times 4$ matrix and $M^D$
a $5\times 5$ one. We will denote the mass eigenstates obtained after
the diagonalization of the matrix in Eq.~(\ref{mass1}) $u,c,t,t'$ and
$d,s,b,d',s'$ for the charge $\frac{2}{3}$ and $-\frac{1}{3}$ sectors
respectively. The primed quarks are new heavy quarks.

Explicitly, we can write the Yukawa interactions for the leptons in
Eq.~(\ref{yu1}) as
\begin{equation}
-{\cal L}_{lY}=G_l(\bar\nu^c_{lL}l_R\eta_1^++\bar\nu_{lL}l_R\eta_2^+
+\bar l_Ll_R\bar\eta^0+H.c.).
\label{ly1}
\end{equation}
We have also the Yukawa term,
\begin{equation}
-{\cal L'}_{lL}=\frac{1}{2}\sum_{l,m}\epsilon^{ijk}h_{lm}\psi_{il}C^{-1}
\psi_{jm}\eta^*_k
\label{ly11}
\end{equation}
with $h_{lm}=-h_{ml}$. Eq.~(\ref{ly11}) implies an antisymmetric
$3\times3$ mass matrix for the neutrinos, and it is well known that
for this kind of matrix one of the eigenvalues is zero and the other
two are degenerated. One way to
obtain an arbitrary mass matrix for the neutrinos, is to introduce a
symmetric sextet $S:({\bf6},+\frac{2}{3})$ or others Higgs multiplets
with the same quantum number of $\eta$ in order to give a mass to the
neutrinos by radiative corrections~\cite{lw,bm}. However,
Eq.~(\ref{ly11}) can be forbidden by introducing an appropriate discrete
symmetry.

The sextet has the following charge assignment
\begin{equation}
S=\left(
\begin{array}{ccc}
S^0_1 & S_2^0 & S^-_1 \\
S^0_2 & S^0_3 & S^-_2 \\
S^-_1 & S^-_2 & S^{--}\end{array}\right).
\label{sextet}
\end{equation}
The $\langle S^0_2\rangle\not=0$ gives a Dirac mass term for
neutrinos. In order to avoid a Majorana mass term we have to chose
$\langle S^0_{1,3}\rangle=0$.  The Yukawa Lagrangian
of the leptons with the sextet is
\begin{equation}
{\cal L}_{lS}=\sum_{lm}G_{lm}\bar \psi^c_{il} \psi_{jm}S^{ij}+H.c.
\label{yl2}
\end{equation}
In Eq.~(\ref{yl2}) $i,j$ denote $SU(3)$ indices and $l,m=e,\mu,\tau$
and $G_{lm}=G_{ml}$. As we have not assigned a lepton number
neither to the $\eta$ nor to the sextet $S$ Higgs multiplets, we have
lepton number violations in Eq.~(\ref{ly1}) and (\ref{yl2}).
\subsection{Models II and III}
\label{subsec:m2y}
As both models have the same quark content and they differ only in
the leptonic sector we will consider them together. We also will
write down the Yukawa interactions for the quarks with charge
$\frac{2}{3},-\frac{1}{3}$. Josions of charge $-\frac{4}{3}$ will mix
to one another but not to the one with charge $\frac{5}{3}$.
Considering only the quarks with charge $\frac{2}{3}$ and
$-\frac{1}{3}$ the quark-Higgs interaction is
\begin{equation}
{\cal L}_Y=\bar Q_{1L}(G^u_{1\alpha}U_{\alpha R}\eta+G^d_{1\alpha}
D_{\alpha R}\rho)+\bar Q_{i\alpha L}(F^u_{i\alpha}U_{\alpha R}\rho^*
+F^d_{i\alpha}D_{\alpha R}\eta^*)+H.c.
\label{yu2}
\end{equation}
where $\alpha=1,2,3$, $i=2,3$, and $U_{\alpha R}=u_{1R},u_{2R},u_{3R}$,
$D_{\alpha R}=d_{1R},d_{2R},d_{3R}$. The mass terms from
Eq.~(\ref{yu2}) are
\begin{equation}
-\bar U_{\alpha' L}M^U_{\alpha'\alpha}U_{\alpha R}
-\bar D_{\alpha' L}M^D_{\alpha'\alpha}D_{\alpha R}+H.c.
\label{mass2}
\end{equation}
In Eq.~(\ref{mass2}) $U_{\alpha'L}=u_{1L},u_{2L},u_{3L}$,
$D_{\alpha' L}=d_{1L},d_{2L},d_{3L}$. $M^U$ and $M^D$ are arbitrary
$3\times 3$ matrices. For the quantum numbers of the
Higgs multiplet $\eta$ and $\rho$ see Ref.\cite{pp}.
\section{Gauge Bosons}
\label{sec:gb}
In this section, we will only consider the model I. For
models II and III which are similar in the gauge sector, see
Ref.~\cite{pp}.

The gauge bosons of this theory consist of an octet $W^a_\mu$
associated with $SU(3)_L$ and a singlet $B_\mu$ associated with
$U(1)_N$. Considering only the triplets of Higgs bosons, the
covariant derivatives are:
\begin{equation}
{\cal D}_\mu\varphi_j=\partial_\mu\varphi_j+ig(\vec
W_\mu\cdot\frac{\vec\lambda}{2})_j^k\varphi_k
+i\frac{g'}{2}N_\varphi B_\mu\varphi_j,
\label{dc}
\end{equation}
where $N_\varphi$ denotes the $N$ charge for the $\varphi$ Higgs
multiplet, $\varphi=\eta,\rho,\sigma$.

We will denote $v_\rho=\langle\rho_1^0\rangle$,
$v_\sigma=\langle\sigma_2^0\rangle$ and $v_\eta=\langle\eta^0\rangle$
the vacuum expectation values of the corresponding Higgs multiplet.
Using them in Eq.~(~\ref{dc}) we obtain the symmetry breaking pattern
\begin{equation}
\begin{array}{c}
SU(3)_L \otimes U(1)_N \\
\downarrow \langle \sigma \rangle\\
SU(2)_L \otimes U(1)_Y\\
\downarrow \langle x\rangle \\
U(1)_{em}
\end{array}
\label{sb}
\end{equation}
where $\langle x\rangle=\langle \eta\rangle,\langle
\rho\rangle$ and the $SU(3)_c$ of color remaining unbroken.

The non-Hermitian gauge bosons $\sqrt2W^+\equiv W^1-iW^2$,
$\sqrt2V^+\equiv W^4-iW^5$ and $\sqrt2X^0\equiv W^6-iW^7$ have the
following masses:
\begin{equation}
M^2_W=\frac{1}{4}g^2\left(v^2_\eta+v^2_\rho\right)\,;
M^2_V=\frac{1}{4}g^2\left(v^2_\eta+v^2_\sigma\right)\,;
M^2_X=\frac{1}{4}g^2\left(v^2_\rho+v^2_\sigma\right).
\label{mq}
\end{equation}
Notice that even if $v_\eta=v_\rho\approx v/\sqrt2$, where $v$ is the
usual vacuum expectation value for the Higgs in the Standard Model,
the VEV $v_\sigma$ must be large enough in order to leave the new
gauge bosons sufficiently heavy to keep consistency with low energy
phenomenology.

Notice that there are two gauge bosons forming an $SU(2)_L$-doublet,
\begin{equation}
\left( \begin{array}{c} V^+\\ X^0\end{array}\right),\quad
\left( \begin{array}{c} \bar X^0\\ V^-\end{array}
 \right),
\label{bx}
\end{equation}
consisting of a charged gauge boson $V^+$ and a neutral one $X^0$.

On the other hand, the neutral (Hermitian)
gauge bosons have the mass matrix $\frac{1}{4}g^2M^2$ in the $(W^3,W^8,B)$
basis, where $M^2$ is given by
\begin{equation}
\left(\begin{array}{ccc}
v^2_\eta+v^2_\sigma & \frac{1}{\sqrt3}(v^2_\eta-v^2_\sigma) &
-\frac{2}{3}t(2v^2_\eta+v^2_\sigma) \\
\frac{1}{\sqrt3}(v^2_\eta-v^2_\sigma) &
\frac{1}{3}(v^2_\eta+v^2_\sigma+4v^2_\rho)
 & -\frac{2}{3\sqrt3}t(2v^2_\eta-v^2_\sigma+2v^2_\rho) \\
-\frac{2}{3}t(2v^2_\eta+v^2_\sigma) &
-\frac{2}{3\sqrt3}t(2v^2_\eta-v^2_\sigma+2v^2_\rho) &
\frac{4}{9}t^2(4v^2_\eta+v^2_\sigma+v^2_\rho)
\end{array}\right),
\label{mn}
\end{equation}
with $t= g'/g$. Since $det M^2=0$ we must have a
photon after the symmetry breaking~\cite{math}. The introduction of the sextet
$S$  spoils the fact that the determinant of Eq.~(\ref{mn})
vanishes. However this can be achieved by imposing a fine
tune on the VEV of the sextet $\langle S^0_2\rangle$ with the VEV's
of the triplets. In fact, this implies that $\langle S^0_2\rangle$ is
of the same order of magnitude of $v_\sigma$ which is the highest VEV
in the model. This is not phenomenologically interesting nor natural since
$\lan
   gle
S^0_2\rangle$ will give mass to the neutrinos. For this reason we
will not introduce the sextet and, since the interaction in
Eq.~(\ref{ly11}) will be forbidden by a discrete symmetry (see
Sec.~\ref{sec:gb}), in the present model the neutrinos remain
massless at tree level. Lepton number violating mass terms can arise
from calculable radiative corrections mediated by gauge or scalar
bosons but, here we will not consider this issue~\cite{vs,lw,bm}.
We must stress that in model II the introduction of a sextet
$({\bf6},0)$ does not spoil the fact that $det \,M^2=0$ without any
restriction on the VEV's~\cite{fhpp}.

The eigenvalues of the matrix in Eq.~(\ref{mn}) are:
\begin{equation}
M^2_A=0,\qquad
M^2_{Z}\simeq
\frac{g^2}{4}\frac{3+4t^2}{3+t^2}(v^2_\eta+v^2_\rho),\qquad
M^2_{Z'}\simeq\frac{g^2}{3}(1+\frac{1}{3}t^2)v^2_\sigma,
\label{auto}
\end{equation}
where we have used $v_\sigma\gg v_{\rho,\eta}$ for the case of $M_Z$
and $M_{Z'}$. Then, we have in this approximation
\begin{equation}
\frac{M^2_Z}{M^2_W}=\frac{3+4t^2}{3+t^2}.
\label{14}
\end{equation}
In order to obtain the usual relation
$\cos^2\theta_W M^2_Z=M^2_W$, we
must have
\begin{equation}
t^2=\frac{s^2_W}{1-\frac{4}{3}s^2_W},
\label{aw}
\end{equation}
where $s^2_W\equiv\sin^2\theta_W$. Hence, we can identified $Z^0$ as the
neutral gauge boson of the Standard Model.

The neutral physical states in the $(W^3,W^8,B)$ basis are:
\[A_\mu=\frac{1}{(3+4t^2)^{\frac{1}{2}}}(\sqrt3t,t,\sqrt3)\]
\[Z_\mu\simeq \frac{-1}{(3+t^2)^{\frac{3}{2}}(3+4t^2)^{\frac{1}{2}}}
\left(\frac{9-3t^2-2t^4}{2},
\frac{27+27t^2+6t^4}{2\sqrt3},-3t(3+t^2)\right) \]
\begin{equation}
Z'_\mu\simeq\frac{1}{(3+t^2)^{\frac{1}{2}}}\left(-\frac{3}{2},
\frac{3}{2\sqrt3}, t\right).
\label{vector}
\end{equation}
\section{Charged and neutral currents}
\label{sec:currents}
The interactions among the gauge bosons and fermions are read off from
\begin{equation}
{\cal L}_F=\bar Ri\gamma^\mu(\partial_\mu+\frac{ig'}{2}B_\mu N_R)R+
\bar Li\gamma^\mu(\partial_\mu+\frac{ig'}{2}B_\mu
N_L+\frac{ig}{2}\vec\lambda\cdot \vec W_\mu)L,
\label{16}
\end{equation}
where $R$ represents any right-handed singlet and $L$ any left-handed
triplets or antitriplets and $N_R(N_L)$ is the $U(1)_N$ charge of the
right-handed (left-handed) fermions. Here also we will consider only
the model I.
\subsection{charged currents}
\label{subsec:cc}
Concerning the charged leptons, we obtain the
electromagnetic interaction by identifying the electric charge $e$ by
\begin{equation}
\vert e\vert=g\sin\theta_W.
\label{e}
\end{equation}
Concerning the interactions with the charged vector fields we have
\begin{equation}
{\cal L}_l=-\frac{g}{\sqrt2}\left(-\bar \nu^c_{lL}\gamma^\mu l_L V^+_\mu
+\bar\nu_{lL}\gamma^\mu l_LW^+_\mu\right)+H.c.,
\label{18}
\end{equation}
where $l=e,\mu,\tau$. There are also currents coupled to the
non-hermitian neutral boson, $\bar\nu^c_{lL}\gamma^\mu
\nu_{lL}X^0_\mu$.

For the first two quark ``generations'' we have,
\begin{equation}
{\cal L}_{Q_{1,2}} = -\frac{g}{\sqrt{2}}\left(
{\bar U}_{iL} \gamma^\mu D_{iL} W^+_\mu -
{\bar U}_{iL} \gamma^\mu D'_{iL} V^+_\mu \right)+H.c.,
\label{cc12}
\end{equation}
where $i=1,2$, $U_i=u_1,u_2$, $D_i\,(D'_i)=d_1,d_2\,(d'_1,d'_2)$.
For the third quark ``generation'' we have
\begin{equation}
{\cal L}_{Q_3} = -\frac{g}{\sqrt{2}}\left(
-\bar u_{3L} \gamma^\mu d_{3L} V^+_\mu +
\bar u_{4L} \gamma^\mu d_{3L} W^+_\mu \right)+H.c.
\label{cc3}
\end{equation}
As $D'_i$ and $u_3$ are $SU(2)_L$ singlet the interaction with $V^+$
violates the lepton number (see Eq.~(\ref{18})) and the weak isospin
(see Eq.~(\ref{cc12})). In this case the
interactions with the neutral boson  $X^0$ are proportional to
\[
-(\bar D_{iL} \gamma^\mu D'_{iL}-\bar u_{3L} \gamma^\mu u_{4L} )X^0_\mu.
\]
Notice that the interactions with the $X^0$ boson also violate the lepton
number and weak isospin. Recall that we have not attributed
lepton number to the gauge bosons, however, as can be seen from
Eq.~(\ref{mq}) the masses of the vector  bosons $V^+,X^0$ are
proportional to the VEV of the $\sigma$, and therefore should be
heavier than $W^+,Z^0$.
\subsection{Neutral currents}
\label{subsec:nn}
Similarly, we have the neutral currents coupled to both $Z^0$ and
$Z'^0$ massive vector bosons. For neutrinos
\begin{equation}
{\cal L}^{NC}_\nu =-\frac{g}{2c_W}\sum_l\left[-\bar\nu^c_{lL}
\gamma^\mu\nu^c_{lL}\left(Z^0_\mu+\frac{1-2s^2_W}{(3-4s^2_W)^{\frac{1}{2}}}
Z'^0_\mu\right)-\frac{2(1-s^2_W)}{(3-4s^2_W)^{\frac{1}{2}}}
\bar\nu_{lL}\gamma^\mu\nu_{lL}Z'^0_\mu\right].
\label{20}
\end{equation}
Using $\bar\nu^c_L\gamma^\mu\nu^c_L=-\bar\nu_R\gamma^\mu\nu_R$, we
see that, in this model, the magnitude of the neutral couplings of
the right-handed neutrinos coincide with that of the left-handed
neutrinos in the Standard Model. In the last equation neutrinos are
still symmetry eigenstates.

Concerning the charged leptons we have
\begin{eqnarray}
{\cal L}^{NC}_l&=&-\frac{g}{2c_W} \sum_l\left(-\bar l_L\gamma^\mu
l_L\left[(1-2s^2_W)Z_\mu+
\frac{1-2s^2_W}{(3-4s^2_W)^{\frac{1}{2}}}Z'^0_\mu\right]\right.
\nonumber \\ && \mbox{}
\left.+2s^2_W\bar l_R\gamma^\mu
l_R[-Z^0_\mu+\frac{1}{(3-4s^2_W)^{\frac{1}{2}}}Z'^0]\right).
\label{ncl}
\end{eqnarray}

Next, let us consider the quark sector. The electromagnetic current for
quarks is the usual one
\begin{equation}
Q_qe\bar q\gamma^\mu qA_\mu,
\label{emc}
\end{equation}
where $q$ is any of the quarks with $Q_q=+\frac{2}{3},-\frac{1}{3}$
and $e$ was defined in Eq.~(\ref{e}).

The neutral currents coupled to the $Z^0$ boson are
$$
-\frac{e}{6}\frac{1}{c_Ws_W}(3-4s^2_W)\bar U_{iL}\gamma^\mu U_{iL}
+\frac{2}{3}e\tan\theta_W\bar U_{iR}\gamma^\mu U_{iR}\,;
$$
$$
-\frac{e}{3}\tan\theta_W\bar D_{iL}\gamma^\mu D_{iL}-
\frac{e}{3}\tan\theta_W\bar D_{iR}\gamma^\mu D_{iR}\,;
$$
$$
\frac{e}{6}\frac{1}{c_Ws_W}(3-2s^2_W)\bar D'_{iL}\gamma^\mu D'_{iL}-
\frac{1}{3}e\tan\theta_W\bar D'_{iR}\gamma^\mu D'_{iR}\,;
$$
$$
-\frac{e}{6}\frac{1}{c_Ws_W}(3-4s^2_W)\bar u_{3L}\gamma^\mu u_{3L}+
\frac{2}{3}e\tan\theta_W\bar u_{3R}\gamma^\mu u_{3R}\,;
$$
$$
\frac{2}{3}e\tan\theta_W\bar u_{4L}\gamma^\mu u_{4L}
+\frac{2}{3}e\tan\theta_W\bar u_{4R}\gamma^\mu u_{4R}\,;
$$
\begin{equation}
\frac{e}{6}\frac{1}{c_Ws_W}(3-2s^2_W)\bar d_{3L}\gamma^\mu d_{3L}
-\frac{1}{3}e\tan\theta_W\bar d_{3R}\gamma^\mu d_{3R}.
\label{nc3}
\end{equation}
The interactions with the $Z'^0$ neutral boson are via the following currents
$$
\frac{e}{6}\frac{(3-4s^2_W)^{\frac{1}{2}}}{c_Ws_W}\bar
U_{iL}\gamma^\mu U_{iL}
-\frac{2}{3}e\frac{\tan\theta_W}{(3-4s^2_W)^{\frac{1}{2}}}\bar
U_{iR}\gamma^\mu U_{iR}\,;
$$
$$
-\frac{e}{3}\frac{(3-4s^2_W)^{\frac{1}{2}}}{c_Ws_W}\bar
D_{iL}\gamma^\mu
D_{iL}+\frac{e}{3}\frac{\tan\theta_W}{(3-4s^2_W)^{\frac{1}{2}}}\bar
D_{iR}\gamma^\mu D_{iR}\,;
$$
$$
\frac{e}{6}\frac{(3-4s^2_W)^{\frac{1}{2}}}{c_Ws_W}\bar
D'_{iL}\gamma^\mu D'_{iL}+
\frac{e}{3}\frac{\tan\theta_W}{(3-4s^2_W)^{\frac{1}{2}}}\bar
D'_{iR}\gamma^\mu D'_{iR}\,;
$$
$$
-\frac{e}{6c_Ws_W}
\frac{(3-2s^2_W)}{(3-4s^2_W)^{\frac{1}{2}}}\bar u_{3L}\gamma^\mu u_{3L}
-\frac{2}{3}e\frac{\tan\theta_W}{(3-4s^2_W)^{\frac{1}{2}}}\bar
u_{3R}\gamma^\mu u_{3R}\,;
$$
$$
\frac{e}{3c_Ws_W}\frac{(3-5s^2_W)}{(3-4s^2_W)^{\frac{1}{2}}}\bar
u_{4L}\gamma^\mu
u_{4L}-\frac{2}{3}e\frac{\tan\theta_W}{(3-4s^2_W)^{\frac{1}{2}}}
\bar u_{4R}\gamma^\mu u_{4R}\,;
$$
\begin{equation}
-\frac{e}{6c_Ws_W}\frac{(3-2s^2_W)}{(3-4s^2_W)^{\frac{1}{2}}}\bar
d_{3L}\gamma^\mu
d_{3L}+\frac{1}{3}e\frac{\tan\theta_W}{(3-4s^2_W)^{\frac{1}{2}}}\bar
d_{3R}\gamma^\mu d_{3R}.
\label{ncz3}
\end{equation}
Here $U=u_1,u_2$, $D=d_1,d_2$ and $D'=d'_1,d'_2$.
\section{GIM Mechanism}
\label{sec:gim}
In this section we will study the GIM mechanism in the models we
have considered before.
In the gauge sector, the neutral currents have the form showed in
Eq.~(\ref{nc}) or
\begin{equation}
-\frac{g}{2c_W}[a_{L}(f)\bar f\gamma^\mu(1-\gamma_5)f+a_{R}(f)\bar
f\gamma^\mu(1+\gamma_5)f]Z^0_\mu,
\label{ncr}
\end{equation}
 for the $Z^0$, and
 \begin{equation}
-\frac{g}{2c_W}[a'_{L}(f)\bar f\gamma^\mu(1-\gamma_5)f+
a'_{R}(f)\bar f\gamma^\mu(1+\gamma_5)f]Z'^0_\mu,
\label{ncr2}
\end{equation}
for the $Z'^0$, where $f$ denotes any fermion.
\subsection{Model I}
\label{subsec:gimi}
For leptons, the neutral currents
appear in Eqs.~(\ref{20}) and (\ref{ncl}).
\begin{equation}
a_L(\nu_l)=0,\quad a_R(\nu_l)=1,
\label{nnnuz}
\end{equation}
In the Standard Model one has $a_L(\nu_l)=1,\,a_R(\nu_l)=0$. For the
couplings of neutrinos with $Z'^0$ we have
\begin{equation}
a'_L(\nu_l)= 0,\quad a'_R(\nu_l)=2\frac{1-x}{g(x)}.
\label{nnnuzp}
\end{equation}
With $l=e,\mu,\tau$. The couplings of the charged leptons with the
$Z^0$ are
\begin{equation}
a_L(l)=-(1-2x),\quad a_R(l)=2x,
\label{nnlz}
\end{equation}
and those with the $Z'^0$,
\begin{equation}
a'_L(l)= \frac{1-2x}{g(x)},\quad a'_R(l)=2x\frac{1}{g(x)}.
\label{nnnuzp2}
\end{equation}
Where we have defined $x=\sin^2\theta_W$ and
$g(x)=(3-4x)^{\frac{1}{2}}$.
Even if neutrinos were massive we can see from Eqs.~(\ref{20}) and
(\ref{ncl}) (or (\ref{nnnuz}) and (\ref{nnlz})) that there is no FCNC
at tree level in the leptonic sector since all lepton representations
transform similarly. This is the case also in models II and III if we
had introduced right-handed neutrinos.

For quarks, recall that the fields are still symmetry eigenstates, we
have
\[
a_L(u_1)=a_L(u_2)=a_L(u_3)=\frac{1}{6}g^2(x),\quad
a_R(u_1)=a_R(u_2)=a_R(u_3)=-\frac{2}{3}x,
\]
\[
a_L(d_1)=a_L(d_2)=\frac{1}{3}x, \quad a_R(d_1)=a_R(d_2)=\frac{1}{3}x,
\]
\[
a_L(d'_1)=a_L(d'_2)=a_L(d_3)=-\frac{1}{6}(3-2x),\quad
a_R(d'_1)=a_R(d'_2)=a_R(d_3)=\frac{1}{3}x.
\]
\begin{equation}
a_L(u_4)=-\frac{2}{3}x,\quad a_R(u_4)=-\frac{2}{3}x.
\label{aqz}
\end{equation}
Similarly for the case of the $Z'^0$ we have
\[a'_L(u_1)=a'_L(u_2)=-\frac{1}{6}g(x),\quad
a'_R(u_1)=a'_R(u_2)=\frac{2}{3}\frac{x}{g(x)},\]
\[
a'_L(d_1)=a'_L(d_2)=\frac{1}{3}g(x),\quad
a'_R(d_1)=a'_R(d_2)=-\frac{1}{3}\frac{x}{g(x)},
\]
\[
a'_L(d'_1)=a_L(d'_2)=-\frac{1}{6}g(x),\quad
a'_R(d'_1)=a_R(d'_2)=-\frac{1}{3}\frac{x}{g(x)},
\]
\[
a'_L(u_3)=\frac{1}{6}\frac{3-2x}{g(x)},\quad a'_R(u_3)=
\frac{2}{3}\frac{x}{g(x)},
\]
\[
a'_L(u_4)=-\frac{1}{3}\frac{3-5x}{g(x)},\quad
a'_R(u_4)=\frac{2}{3}\frac{x}{g(x)},
\]
\begin{equation}
a'_L(d_3)=\frac{1}{3}\frac{3-2x}{g(x)},\quad
a'_R(d_3)=-\frac{1}{3}\frac{x}{g(x)}.
\label{aqzp}
\end{equation}
Notice that as quarks of the same charge gain mass from different
sources, in Eqs.~(\ref{yu1}) we have FCNC at tree level. In fact, the
quark mass matrices in Eq.~(\ref{mass1}) arising from (\ref{yu1}) are
$4\times 4$ and $5\times 5$ for the charge $\frac{2}{3}$ and
$-\frac{1}{3}$ respectively. After diagonalizing these matrices we
obtain mixing among all quarks of the same charge. Since the coupling
appearing in Eq.~(\ref{aqz}) do not satisfy the condition (\ref{i})
the GIM mechanism does not work. One way, but probably not the only
one, to overcome this trouble is the introduction of appropriate
discrete symmetries. We can see from (\ref{aqz}) that if the quarks
$u_1,u_2,u_3$ do not couple to the quark $u_4$, and the quarks
$d'_1,d'_2,d_3$ do not couple to the quarks $d_1,d_2$ the mass matrix
will have the form
\begin{equation}
\left(\begin{array}{cc}
M_1&0\\0&M_2\end{array}\right)\,\mbox{for the charge $-\frac{1}{3}$ sector},
\label{m1}
\end{equation}
\begin{equation}
\left(\begin{array}{cc}
M_3&0\\0&1\end{array}\right)\,\mbox{for the charge $\frac{2}{3}$ sector},
\label{m2}
\end{equation}
where $M_1,M_3$ are $3\times 3$ matrices and $M_2$ a $2\times 2$ matrix. To
constrain the quark mass matrix to have the form given in
Eqs.~(\ref{m1}) and (\ref{m2}) we impose the following discrete
symmetries on the Yukawa Lagrangian of Eq.(\ref{yu1}).
\begin{equation}
\begin{array}{c}
(\eta,\sigma)\to (\eta,\sigma)\\
\rho \to -\rho\\
(\psi_{lL},R_l)\to (\psi_{lL},R_l)\\
(u_{1R},u_{2R},u_{3R})\to (u_{1R},u_{2R},u_{3R})\\
u_{4R}\to -u_{4R}\\
(d'_{1R},d'_{2R},d_{3R})\to (d'_{1R},d'_{2R},d_{3R})\\
(d_{1R},d_{2R})\to -(d_{1R},d_{2R}).
\end{array}
\label{ds}
\end{equation}
Taken into account these symmetries we can rewrite the quark sector
of Eq.~(\ref{yu1}) as follows:
\begin{eqnarray}
-{\cal L}_{QY}&=&\bar Q_{iL}G^u_{i\alpha}U_{\alpha R}\eta+\bar
Q_{3L}G^b_{3\alpha}D_{\alpha R}\eta\nonumber \\ & &\mbox{}
+\bar Q_{iL}\left(G^d_{i\alpha}D_{\alpha R}\sigma+G'^d_{ia}D_{aR}\rho\right)
\nonumber \\ & &\mbox{}
+\bar Q_{3L}\left(G'^t_{3\alpha}U_{\alpha R}\sigma^*+u_{4R}\rho^*\right)+H.c.,
\label{yu11}
\end{eqnarray}
where $i=1,2$, $\alpha=1,2,3$, $a=4,5$ and $U_{\alpha
R}=u_{1R},u_{2R},u_{3R}$, $D_{\alpha R}=d'_{1R},d'_{2R},d_{3R}$ and
$D_{aR}=d_{1R},d_{2R}$. The mass matrices have the tensor product
form in (\ref{m1}) and (\ref{m2}) next, they can be diagonalized with
unitary matrices which are themselves tensor product of unitary
matrices.

We see that, by imposing the symmetries (\ref{ds}), we have the
GIM mechanism in the neutral currents coupled to the $Z^0$,
separately, in the following sectors: $(u_1,u_2,u_3)$; $(d_1,d_2)$
and $(d'_1,d'_2,d_3)$. The $u_4$ quark does not mix at all. Notice
that it is the sector $(u_1,u_2,u_3)$ which have the same neutral
couplings than the $u,c,t$ quarks of the Standard Model. The same
occurs for the $(d'_1,d'_2,d_3)$ which have the same neutral
couplings than the $d,s,b$ of the Standard Model. Notice also that
$d_1,d_2$ have only vector currents coupled to $Z^0$.

On the other hand, there are FCNC in the scalar sector since quarks
with the same charge still gain mass from different Higgs fields. The
Yukawa couplings of the charge $\frac{2}{3}$
quarks $(u_1,u_2,u_3)$ to the neutral Higgs $\eta^0$ and $\sigma^0$,
defined as
\[\eta^0=v_\eta+\zeta\equiv v_\eta+\zeta_1+i\zeta_2 \]
and
\[\sigma^0=v_\sigma+\zeta'\equiv v_\sigma+\zeta'_1+i\zeta'_2,\]
can be written as
\begin{equation}
(1+\frac{\zeta}{v_\eta})\bar
U_{\alpha}M^u_{\alpha\alpha'}U_{\alpha'}+
\left(\zeta'-\frac{v_\sigma}{v_\eta}\zeta\right)\bar
u_{3L}G'^t_{3\alpha}U_{\alpha R}.
\label{u}
\end{equation}
We recall that all fields in Eq.~(\ref{u}) are still phenomenological
states, that is, linear combinations of the mass eigenstates. We will
denote mass eigenstates $u,c,t'$ in the $u_1,u_2,u_3$ sector. Then,
from eq.~(\ref{u}) we see that we will have $u\leftrightarrow c$
transitions.

This situation also occurs in the  charge $-\frac{1}{3}$ sector composed by
$d'_1,d'_2,d_3$ (with the respective mass eigenstates denoted $d',s',b$):
\begin{equation}
(1+\frac{\zeta'}{v_\sigma})\bar
D_{\alpha}M^d_{\alpha\alpha'}D_{\alpha'}+
\left(\zeta-\frac{v_\eta}{v_\sigma}\zeta'\right)\bar
d_{3L}G^b_{3\alpha}D_{\alpha R}.
\label{u2}
\end{equation}
We see that there are flavor changes in both charge sectors induced by the
physical scalars.

However, we must notice that as the fields $\zeta_{1,2}$ and
$\zeta'_{1,2}$ are linear combinations of mass eigenstates, say
$\zeta_i=V_{ij}\phi_j$ and similarly with the $\zeta'_i$ fields.
Next, suppose we choose one of the mass eigenstates $\phi_1$. From
Eq.~(\ref{u2}) we will obtain terms like the following
\begin{equation}
(V_{i1}-\frac{v_\sigma}{v_\eta}V'_{i1})V_{bd'}\bar b_Ld'_R\phi_1+H.c.,
\label{eff}
\end{equation}
with $i$ fixed. In the last equation all fields are already mass
eigenstates, the matrix $V_{bd'}$ is a typical set of mixing angles among
the charge $-\frac{1}{3}$ sector $(d',s',b)$.

We see that interactions in Eq.~(\ref{eff}) produce FCNC in the
$(d',s',b)$ system, but no one of them produce the transition
appearing in Eqs.(\ref{exp1})-(\ref{exp5}). In the charge
$\frac{2}{3}$ sector there are also interactions like those in
Eq.(\ref{eff}). Anyway, these interactions do not imply strong
constraints on the masses of the neutral scalars if conditions like
$V_{i1}-\frac{v_\sigma}{v_\eta}V'_{i1}\simeq0$ are valid.

The quarks $d_1,d_2$ have the GIM mechanism in the coupling with
$Z^0$, $Z'^0$ and also in the scalar sector. The respective mass
eigenstate are denoted by $d,s$.
 \subsection{Models II and III}
\label{subsec:gimii}
With respect to model II the respective coupling are given in
Ref.~\cite{pp}. In this case we have the same couplings for each
charge sector, that is, we have the relation in Eq.~(\ref{y}) valid
and when we redefine the quarks fields as in Eq.~(\ref{def}). FCNC
appear with the coupling to the $Z'^0$ but not with the $Z^0$. The
same for the case of model III. Here we will rewrite the coefficients
calculated in Ref.~\cite{pp}.
Couplings with $Z^0$:
\[a_L(\nu)=1,\quad a_R(\nu)=0,\]
\[a_L(l)=-1+2x,\quad a_R(l)=2x,\]
\[a_L(u_1)=a_L(u_2)=a_L(u_3)=\frac{1}{6}g^2(x), \,
a_R(u_1)=a_R(u_2)=a_R(u_3)=-\frac{2}{3}x,\]
\begin{equation}
a_L(d_1)=a_L(d_2)=a_L(d_3)=-\frac{1}{6} (3-2x) ,\;
a_R(d)=a_R(s)=a_R(b)=\frac{1}{3}x.
\label{nciiz}
\end{equation}
Couplings with $Z'^0$:
\[a'_L(\nu_l)=-\frac{1}{2\sqrt3}h(x),\quad a'_R(\nu_l)=0,\]
\[a'_L(l)=-\frac{h(x)}{2\sqrt3}, \quad a'_R(l)=-\frac{1}{\sqrt3}h(x),\]
\[a'_L(u_1)=-\frac{1}{2\sqrt{3}h(x)},\quad
a'_R(u_1)=-\frac{2}{\sqrt3}\frac{x}{h(x)}\]
\[a'_L(d_1)=-\frac{1}{2\sqrt{3}h(x)},\quad
a'_R(d_1)=\frac{2x}{2\sqrt{3}h(x)},\]
\[a'_L(u_2)=a'_L(u_3)=\frac{1}{4\sqrt{3}}h(x),\quad
a'_R(u_2)=a'_R(u_3)=-\frac{2}{\sqrt{3}h(x)}x,\]
\begin{equation}
a'_L(d_2)=a'_L(d_3)=\frac{1-2x}{2\sqrt{3}h(x)}\quad
a'_R(d_2)=a'_R(d_3)= \frac{x}{\sqrt{3}h(x)},
\label{nciizp}
\end{equation}
where $h(x)=(1-4x)^{\frac{1}{2}}$.

We see from Eqs.~(\ref{nciiz}) and (\ref{nciizp}) that there is GIM
mechanism in the couplings with $Z^0$ but not in the couplings with
$Z'^0$ as it was stressed in Ref.\cite{pp}. It is also necessary in
model II to introduce discrete symmetries if we want to prevent
neutrinos gain mass but we will not consider
this in this paper~\cite{pp2}.

As in the case of model I (see Eq.~(\ref{u})), we can verify that
there are scalars coupled to $\bar u_RG^u_{1\alpha}U_{\alpha R}$
implying mixing among $u,c,t$ and also scalars coupled to
$\bar d_LG^d_{1\alpha}D_{\alpha  R}$ producing mixing among $d,s,b$,
but as it will also appear factors like
$V_{i1}-\frac{v_\sigma}{v_\eta}V'_{i1}$, this does not impose
necessarily a strong lower bound on the neutral scalar masses.

\section{Conclusions}
\label{sec:con}
Following the generalization of the GIM mechanism of Ref.~\cite{gp},
we have considered three different representation contents in a
theory of electroweak interactions with $SU(3)_L\otimes U(1)_N$
gauge symmetry and we have studied in detail the neutral currents.
However, it should be noticed that in all these models we have
considered, fields of the same charge belong to different
representations and for this reason the mixings are not neither pure
``horizontal'' nor pure ``vertical'', in the sense of Ref.~\cite{gp}.
For example there are mixings in the system $d',s',b$ and these mass
eigenstates are linear combinations of two $SU(2)$ singlet,
$d'_1,d'_2$, and one member of an $SU(2)$ doublet, $d_3$. We must
recall that this choice of the representation content was necessary
in order to avoid anomalies in the theory.

Here we will revise and comment the FCNC effects and GIM mechanism in
this kind of models. As in the Standard Model, all neutral couplings
depend on the weak angle $\theta_W$, which can be determined from
several neutral currents processes, the $W^+$ and $Z^0$ masses and
also from $Z$-pole observables like the width and some asymmetries.
It is well known that these experiments have such a level of
precision that complete $O(\alpha)$ radiative corrections are
mandatory. In particular some radiative corrections modify the
tree-level expressions for neutral currents processes.

Let us start by model I. Table~\ref{t1} summarizes the
situation.

Notice that the mass eigenstates $u,c,t'$ and $d',s',b$ have the same
neutral current couplings than the $u,c,t$ and $d,s,b$, respectively,
in the electroweak standard model. However, one quark with charge
$\frac{2}{3}$ quark, say $t'$, and two quarks with charge $-\frac{1}{3}$,
say $d',s'$, can be chosen to have their masses depending
mainly on the $v_\sigma$ which is the VEV that it is the responsible
of the first symmetry breakdown, hence these quarks may be heavier
than the others, as is demanded by experimental data.

However in the charged currents the mixing is not the usual one. For
example, in terms of the mass eigenstates, we have the interactions
with the $W^+$ vector boson (which coincide with the charged vector
boson of the Standard Model),
\begin{equation}
-\frac{g}{\sqrt2}(\bar u\,\bar c\, \bar t)
\left(\begin{array}{cc}
U &0 \\
0 &1\end{array}\right)\gamma^\mu\left(\begin{array}{c}
d\\s\\b_\theta\end{array}\right)W^+_\mu+H.c.,
\label{}
\end{equation}
where $U$ is a $2\times 2$ Cabibbo-like mixing matrix and
$b_\theta=c_1d'+c_2s'+c_3b$, being $c_i$ appropriate
combinations of mixing angles in the $d',s',b$ sector. In the
present model, we have also the charged vector boson $V^+$
\begin{equation}
-\frac{g}{\sqrt2}(\bar u\,\bar c\, \bar t')U'
\gamma^\mu\left(\begin{array}{c}
d'\\s'\\b\end{array}\right)V^+_\mu+H.c.,
\label{67}
\end{equation}
and $U'$ is an arbitrary $3\times 3$ matrix.

Notice that in this model transitions as $u,c\leftrightarrow d,s$ and
$t\leftrightarrow b_\theta$ proceed via $W^+$ exchange. However,
those as $u,c,t'\leftrightarrow d',s',b$ happen via $V^+$ exchange.
It means that semileptonic decays of $B$ mesons (which are very
suppressed) imply a constraint on the $V^+$ mass. In the Standard
Model, the suppression factor, say in $b\rightarrow cl\bar\nu_l$
transitions, is $\vert V_{cb}\vert^2$, where $V_{cb}$ is the respective
Kobayashi-Maskawa matrix element, and in the present model it is
$U'^2_{cb}(M_W/M_V)^4$ , with $U'$ the mixing matrix element
appearing in Eq.~(\ref{67}). This implies, using $\vert
V_{cb}\vert\sim0.0043$~\cite{pdg} that $M_V=4.8M_W\sqrt{U'_{cb}}$. If
$U'_{cb}\sim 1$ the suppression in $b\rightarrow u$ transitions arises
because of $U'_{ub}\sim0.1$. Hence the mass of the $V^+$ field is
of the order $4.8\,M_W\sim384$ GeV. This means from Eqs.~(\ref{mq}),
(\ref{auto}) and (\ref{aw}) that $M_{Z'}> 5.8\,M_W\sim 470$ GeV.

In this model the expression for the couplings $a_L$ and
$a_R$ of quarks $u,c,t'$ and $d',s',b$ coincide with the respective
$a_L$ and $a_R$ of the Standard Model but the quarks $d$ and $s$ have
pure (at tree level) vector neutral currents.

We can verify that this model is not inconsistent with present day
experiments.

As an example, let us consider the deep inelastic neutrino scattering
from approximately isoscalar targets. Experimentally it was measured
the ratio $R_\nu\equiv\sigma^{NC}_{\nu N}/\sigma^{CC}_{\nu N}$ where
$NC$ and $CC$ denote neutral and charged current respectively. In the
simplest approximation
\begin{equation}
R_\nu=g^2_L+g^2_Rr.
\label{Rnu}
\end{equation}
Assuming that there is a contribution from the $d'$ quark besides those of
the $u,d$ and using the constants $a_{L,R}$ defined in
Eq.~(\ref{aqz}), at the zero$^{th}$-order approximation, ones
has~\cite{ua,costa,pl},
\begin{equation}
g^2_L\equiv a^2_L(u)+a^2_L(d')+a^2_L(d)=\frac{1}{2}-x+\frac{2}{3}x^2\simeq0.304
\label{gl}
\end{equation}
\begin{equation}
g^2_R\equiv a^2_R(u)+a^2_R(d')+a^2_R(d)=\frac{2}{3}x^2\simeq0.036,
\label{gr}
\end{equation}
where we have used $x\equiv\sin^2\theta_W\simeq0.232$. In (\ref{Rnu})
$r\equiv\sigma^{CC}_{\bar\nu N}/\sigma^{CC}_{\nu N}$ is the ratio of
$\bar\nu$ and $\nu$ charged current cross section. Assuming the value
of the parton model, $r\simeq0.440$ we obtain that in model I,
$R_\nu\simeq0.320$ which lies in the range of the measured values
from several isoscalar targets~\cite{ua}.

Similarly, it is easy to see that if the only quark that gives an
additional contribution to the
$Z^0$-width is the $d'$, then this model is consistent with experimental data.
The partial width for the $Z^0$ to decay into massless fermions is
$$
\Gamma(Z \to \bar \Psi_i \Psi_i)=\frac{CG_FM_Z^3}{6\sqrt{2}\pi}(V_i^2+A_i^2)
$$
For leptons $C=1$ while for quarks $C=3$ without QCD corrections, and
$V^i=a^i_L + a^i_R$, $A^i=a^i_L - a^i_R$ with $a_L$ and $a_R$
defined in Eq.(~\ref{aqz}). Considering contributions of
leptons and of the quarks $u,c,d,d',s$ and $b$ we obtain
$\Gamma_Z \sim 2.600$ GeV
( which is not too
different from the value in the Standard Model without radiative corrections).
The experimental value is $\Gamma_Z\simeq2.487\pm0.010$
GeV~\cite{pdg}. The small discrepancy could be explained taken into
account radiative corrections and uncertainties coming from QCD. The
forward-backward asymmetry of quark pairs
measured in $e^+e^-$ processes also is sensitive to the weak angle,
but in this case (as in the case of the width) it is necessary to use
in the calculations the effective weak angle at the $Z^0$ mass.
Hence, radiative corrections in this model must be calculated.

Summarizing, we have shown that in model I, by imposing the
discrete symmetry given in  Eq.~(\ref{ds}) we have separated the
quark sector in two charge $\frac{2}{3}$, $u,c,t'$ and $t$, and two
charge $-\frac{1}{3}$ ones, $d',s',b$ and $d,s$. In this situation
the GIM mechanism works in each separate sector in the neutral
currents coupled to the lightest neutral gauge bosons, $Z^0$.
This means that the new quark $d'$ must not
be too heavy since its contribution to the $R_\nu$ and $\Gamma_Z$ are
necessary to fit experimental data.

Another possibility is that the symmetry breaking of the
$SU(3)\otimes U(1)$ symmetry occurs at an energy scale which is not
too different from the Fermi scale. In this case all exotic gauge
bosons are heavier but not too much than the $W^\pm, Z^0$ gauge bosons
of the Standard Model implying a rich phenomenology. However, our
expression are not valid in this situation.

Next, in models II and III, the FCNC in the gauge sector are
restricted mainly to the $Z'^0$ exchange.
See Eqs.(~\ref{nciiz}) and (~\ref{nciizp}). This boson also gains mass
from the VEV which induce the first symmetry breaking. In this case,
however, constraints on the FCNC coming from the $K^0-\bar K^0$
system imply that this energy scale is greater than 8 TeV,
that is, $M_{Z'}>40$ TeV~\cite{pp}. In these
models since there are only new quarks with charge $\frac{5}{3}$,
$-\frac{4}{3}$, the phenomenology of the usual charge sectors, i.e.,
$\frac{2}{3},-\frac{1}{3}$ decouples from that of the exotic sector
and it is automatically consistent with the present observation if the
VEV which is in control of the breaking of the $SU(3)_L\otimes
U(1)_N$ symmetry are fixed by the FCNC contributions of the $Z'^0$
neutral boson.

In the three models, FCNC also appear in the Yukawa
interactions, however, these involve some of the new quarks, therefore
there is no strong constraints coming from experimental data. Anyway,
the suppression factors as in Eq.~(\ref{eff}) implies that the masses of the
neutral Higgs scalars are not necessarily very high in order to have
the appropriate suppression of the FCNC.

The new gauge bosons as $V^+,X^0$ in model I, only couple one of the
light $(u,c,t,b)$ to one of the heavy $(d',s',t')$ quarks, they
cannot be produced in current experiments. FCNC induced by box
diagrams by the $V^+$ boson involve also the heavy quarks and for
this reason it does not imply strong constraints on its mass.
Interfamily lepton number violations provide weaker bounds on these
mass (and on that of the $X^0$ boson) also. The same happens with
Zezeons and Wanions in models II and III~\cite{pp}. Even the neutrino
counting experiment
$e^+e^-\to\nu^c\nu^c\gamma$ is not so restrictive and since the GIM
mechanism works in the leptonic sector even if the neutrinos become
massive, process as $\mu\to e\nu_e^c\nu_\mu^c$ will be very
suppressed. In fact, these processes imply that masses of all the new
gauge bosons of the order of two or three times the mass of the
charged gauge boson in the standard electroweak model.

We think that these models, being consistent with low energy
data and because of the color degree of
freedom and the number of families are related in order to be the
models anomaly free, they could imply unforeseen options
in model building. New physics can arise at not too high energies.
\begin{center}
{\bf Acknowledgments}
\end{center}
We would like to thank the Con\-se\-lho Na\-cio\-nal de
De\-sen\-vol\-vi\-men\-to Cien\-t\'\i \-fi\-co
e Tec\-no\-l\'o\-gi\-co (CNPq) for full (J.C.M.\ ,\ F.P.) and partial
(V.P.) financial support. We also are very gratefully to G.E.A.
Matsas, and one of us (V.P.) to O.F. Hernandez, for useful
discussions.

\unletteredappendix{}
As we have used $\bar\lambda$'s matrices, for ${\bf3}^*$ antitriplet
representation of $SU(3)$ group, which are not usually found in the
literature, we will consider them explicitly in this
Appendix~\cite{laf}.

The up and down operators in the triplet representation  are given by
 \begin{equation}
\begin{array}{ccc}
E_1=\left(\begin{array}{ccc}
0\, & 1\, & 0\, \\ 0\, & 0\, & 0\, \\ 0\, & 0\, & 0\,
\end{array}\right),&
E_2=\left(\begin{array}{ccc}
0\, & 0\, & 0\, \\ 0\, & 0\, & 1\, \\ 0\, & 0\, & 0\,
\end{array}\right),&
E_3=\left(\begin{array}{ccc}
0\, & 0\, & 1\, \\ 0\, & 0\, & 0\, \\ 0\, & 0\, & 0\,
\end{array}\right);
\end{array}
\label{a1}
\end{equation}
\begin{equation}
\begin{array}{ccc}
E_{-1}=\left(\begin{array}{ccc}
0\, & 0\, & 0\, \\ 1\, & 0\, & 0\, \\ 0\, & 0\, & 0\,
\end{array}\right),&
E_{-2}=\left(\begin{array}{ccc}
0\, & 0\, & 0\, \\ 0\, & 0\, & 0\, \\ 0\, & 1\, & 0\,
\end{array}\right),&
E_{-3}=\left(\begin{array}{ccc}
0\, & 0\, & 0\, \\ 0\, & 0\, & 0\, \\ 1\, & 0\, & 0\,
\end{array}\right),
\end{array}
\label{a2}
\end{equation}
with the following commutation relations
\[[E_1,E_3]=[E_1,E_{-2}]=[E_2,E_3]=0;\]
\begin{equation}
[E_1,E_{-3}]=-E_{-2},\quad [E_1,E_2]=E_3,\quad [E_2,E_{-3}]=E_{-1}.
\label{a3}
\end{equation}
The Cartan subalgebra generators are represented by
\begin{equation}
H^{\lambda}_1=\mbox{diag}(1,-1,0);\quad
H^{\lambda}_2=\mbox{diag}(0,1,-1).
\label{a4}
\end{equation}
The $SU(3)$ algebra is generated by real linear combination of the
matrices $H_k$, $(E_l+E_{-l})$ and $i(E_l-E_{-l})$; $k=1,2;
l=1,2,3$. Then the so called Gell-Mann matrices $\lambda$ are:
\[
\lambda_1\equiv E_1+E_{-1}=\left(\begin{array}{cc}
\sigma_1\, & 0\, \\ 0\, & 0\,\end{array}\right),\;
 \lambda_2\equiv -i(E_1-E_{-1})=\left(\begin{array}{cc}
\sigma_2\, & 0\, \\ 0\, & 0\,
\end{array}\right), \]
\[ \lambda_3\equiv H^{\lambda}_1=\left(
\begin{array}{cc}
\sigma_3\, & 0\,\\ 0\,&0\,\end{array}\right),\]

\[
 \lambda_4\equiv E_3+E_{-3}=\left(\begin{array}{ccc}
0\, & 0\, & 1\, \\ 0\, & 0\, & 0\, \\ 1\, & 0\, & 0\,
\end{array}\right), \;
 \lambda_5\equiv -i (E_3-E_{-3})=\left(\begin{array}{ccc}
0\, & 0\, & -i\, \\ 0\, & 0\, & 0\, \\ i\, & 0\, & 0\,
\end{array}\right), \]
\[
\lambda_6\equiv E_2+E_{-2}=\left(\begin{array}{cc}
0\, & 0\, \\ 0\, & \sigma_1\,
\end{array}\right), \;
 \lambda_7\equiv -i (E_2-E_{-2})=\left(\begin{array}{cc}
0\, & 0\, \\ 0\, & \sigma_2\,
\end{array}\right), \]
 \begin{equation}
\lambda_8\equiv\frac{1}{\sqrt3}(H^{\lambda}_1+2H^{\lambda}_2)=
\mbox{diag}(\frac{1}{\sqrt3},\frac{1}{\sqrt3},-\frac{2}{\sqrt3}),
\label{a5}
\end{equation}
where $\sigma_i$ are the usual $2\times 2$ Pauli matrices.
The $\lambda_a, a=1,2,...,8$, have the normalization
\begin{equation}
\mbox{tr}(\lambda_a\lambda_b)=2\delta_{ab},
\label{a7}
\end{equation}
and the closed algebra
\begin{equation}
\left[\frac{\lambda_a}{2},\frac{\lambda_b}{2}\right]=
if_{abc}\frac{\lambda_c}{2}
\label{a8}
\end{equation}
with the following non--vanishing values for the totally antisymmetric
structure constant $f_{abc}$:
\begin{equation}
\begin{array}{l}
f_{123}=1, \\
f_{147}=-f_{156}=f_{246}=f_{257}=f_{345}=-f_{367}=\frac{1}{2}, \\
f_{458}=f_{678}=\frac{\sqrt3}{2}
\end{array}
\label{a9}
\end{equation}
As the $SU(3)$ is a rank 2 group, the diagonal matrices
$\lambda_3$ and $\lambda_8$ are such that
\begin{equation}
\left[\frac{\lambda_3}{2},\frac{\lambda_8}{2}\right]=0,
\label{a10}
\end{equation}
and $\lambda_{1,2,3}$ are a representation of the
subgroup $SU(2)$ algebra.
\par
The other fundamental representation of
$SU(3)$ of dimension 3, {\it i.e.} the anti`triplet representation
$\bar{\lambda}$ can be obtained by the following substitution:
\begin{equation}
\begin{array}{c}
E_{\pm 1}\longrightarrow E_{\pm 2}, \\
E_{\pm 2}\longrightarrow E_{\pm 1}, \\
E_{\pm 3}\longrightarrow -E_{\pm 3},
\end{array}
\label{a11}
\end{equation}
and
\begin{equation}
H^{\bar{\lambda}}_1=H^{\lambda}_2 ,\,\,\,
H^{\bar{\lambda}}_2=H^{\lambda}_1.
\label{a12}
\end{equation}
We have explicitly:
\[
\bar{\lambda_1} = E_2+E_{-2}=\lambda_6, \;
\bar{\lambda_2} = -i (E_2-E_{-2})=\lambda_7,
\]
\[
\bar{\lambda_3} = H^{\bar{\lambda}}_1=\mbox{diag}(0,1,-1),\;
\bar{\lambda_4} = -(E_3+E_{-3})=-\lambda_4,
\]
\[
\bar{\lambda_5} = i (E_3-E_{-3})=-\lambda_5,\;
\bar{\lambda_6} = E_1+E_{-1}=\lambda_1,
\]
\[
\bar{\lambda_7} = -i (E_1-E_{-1})=\lambda_2, \]
\begin{equation}
\bar{\lambda_8} =\frac{1}{\sqrt3}(H^{\bar{\lambda}}_1+2H^{\bar{\lambda}}_2)=
\mbox{diag}(\frac{2}{\sqrt3},-\frac{1}{\sqrt3},-\frac{1}{\sqrt3}).
\label{a13}
\end{equation}
Using Eqs.~(\ref{a13}) in Eq.~(\ref{oc1}) we can verify that the leptons
are in antitriplets and that the Higgs antitriplets are
\begin{equation}
\begin{array}{ccc}
\eta^*=\left(\begin{array}{c}
\eta^+_1 \\ \eta_2^+ \\ \bar\eta^0
\end{array}\right)\sim ({\bf3}^*,\frac{2}{3});&
\rho^*=\left(\begin{array}{c}
\bar\rho_2^0 \\ \bar\rho_1^0 \\ \rho^-
\end{array}\right),&
\sigma^*=\left(\begin{array}{c}
\bar\sigma_2^0 \\ \bar\sigma_1^0 \\ \sigma^-\end{array}\right)\sim
({\bf3}^*,-\frac{1}{3}),
\end{array}
\label{at}
\end{equation}
It is well known that the anomaly is proportional to~\cite{gg}
\[
{\cal A}({\bf3})\propto Tr(\{\lambda_a,\lambda_b\},\lambda_c),
\]
for the triplet, and
\[
{\cal A}({\bf3}^*)\propto Tr(\{\bar\lambda_a,\bar\lambda_b\},\bar\lambda_c),
\]
for the antitriplet representation. Using the $\lambda$'s and
$\bar\lambda$'s given above it is straightforward to verify that
\begin{equation}
{\cal A}({\bf3})=-{\cal A}({\bf3}^*).
\label{last}
\end{equation}

\begin{table}
\caption{FCNC effects in model I. $Z^0$ and $Z'^0$ are the lightest
and the heaviest neutral boson and $H^0$ is a typical neutral
scalar. The ``yes'' and ``no'' in the entries denote the
existence or not of FCNC respectively.}\label{t1}
\begin{tabular}{|c|c|c|c|}
Quark Sector & $Z^0$ & $Z'^0$ & $H^0$ \\ \hline\
$u,c,t'$     &  no   &  yes   & yes \\
$t$          &  no   &  no    & no   \\
$d,s$        &  no   &  no    & no \\
$d',s',b$    &  no   &  yes   & yes
\end{tabular}
\end{table}
\end{document}